\documentclass[aps,prl,floatfix,twocolumn,footinbib]{revtex4}

\usepackage{epsfig}
\usepackage{bm}
\usepackage{mathtools}
\usepackage{color}
\usepackage{framed}
\usepackage{cancel}
\usepackage{tikz}
\usepackage{appendix}

\newcommand\zed{\mathcal{Z}}

\let\oldref\ref
\renewcommand\ref[1]{(\oldref{#1})}

\newcommand\eq[1]{\[#1\]}
\newcommand\tr[1]{\textrm{#1}}

\newcommand\eqn[1]{\begin{equation}#1\end{equation}}
\newcommand\alg[1]{\begin{align*}#1\end{align*}}
\newcommand\algn[1]{\begin{align}#1\end{align}}
\newcommand\de{\textrm{d}}

\begin{document}

\title{Quantum Hall Edges with Hard Confinement:  Exact Solution beyond Luttinger Liquid}
\author{Richard Fern}
\affiliation{The Rudolf Peierls Centre for Theoretical Physics, Oxford University, Oxford OX1 3NP}
\author{Steven H. Simon}
\affiliation{The Rudolf Peierls Centre for Theoretical Physics, Oxford University, Oxford OX1 3NP}

\begin{abstract}
We consider a Laughlin droplet in a confining potential which is very steep but also weak compared to the ultra-short ranged inter-particle interactions.  We find that the eigenstates have a Jack polynomial structure, and have an energy spectrum which is extremely different from the well-known Luttinger liquid edge.   
\end{abstract}

\maketitle

The edges of quantum Hall (QH) systems, a paradigm for the protected edge modes of topological matter, have been a topic of intense interest for over two decades\cite{Chang}.
Classic work\cite{WenEdge} describes the QH edge as a chiral Luttinger liquid with excitation frequency proportional to wavevector, and extensive experiments have supported this picture\cite{Chang}.
More recently a number of theoretical works have examined deviations from the linear spectrum\cite{PriceQHD, CooperSignatures, ImambekovGlazmann, WiegmannNonlinear, BettelheimShocks}.
One line of thought suggests more general edge hydrodynamics\cite{WiegmannNonlinear, BettelheimShocks} related to a quantum version of the Calogero-Sutherland model\cite{ZNCHa,DiejenCalogero,KuramotoCalogero}, although microscopic derivations of these ideas has been lacking in the quantum Hall context.  

In the current work we study the  edge states of Laughlin quantum Hall droplets in a solvable limit with almost hard-wall confinement and ultra-short range interactions.
We find a spectrum which is extremely different from the traditional chiral Luttinger liquid picture --- where excitation branches have minimum energy at high angular momentum.
The exact eigenstates are found to have a Jack polynomial structure analogous to a corresponding Calogero-Sutherland model\cite{ZNCHa}, however the eigenenergies bear no resemblance to that model.
While Jack polynomials have been used before in the quantum Hall context\cite{BernevigJacks, BernevigAnatomy, Estienne, Gurarie}, we emphasise that the structure found here is quite different from that discussed previously.

We work in the lowest Landau level and consider the special interaction $\hat V_m$ which gives positive energy to any two particles with relative angular momentum less than $m$  (and zero energy for higher relative angular momenta).
The Laughlin $\nu=1/m$ droplet is the unique highest density zero energy state of this interaction (assuming bosonic particles for $m$ even and fermionic for $m$ odd).
Using symmetric gauge we write the Laughlin droplet  as
	\eq{\Psi_{m,N} = \zed_N^{-1/2} \mbox{$\prod_{i<j=1}^N$} (z_i-z_j)^m  e^{-\sum_{i=1}^N|z_i|^2/4},}
where $z_i=x_i+iy_i$ is the complex position of the $i^\tr{th}$ particle, the magnetic length is set to unity, and $\zed_N$ is the $N$-particle normalisation of the wavefunction, defined by $\langle \Psi_{m,N} | \Psi_{m,N} \rangle  = 1$.
For this special interaction, any symmetric polynomial in the $z_i$'s multiplied by $\Psi_{m,N}$ is also zero energy for this interaction, representing some edge excitation.
A homogeneous polynomial of total degree $\Delta L$ would correspond to adding a total angular momentum $\Delta L$ to the droplet (because $z_i^l$ refers to a single-particle orbital at angular momentum $l$).

We now consider a rotationally symmetric confining potential $\hat U$ added to the Hamiltonian which we can generally write as 
$\hat U = \sum_lU_l\hat n_l$, where $\hat n_l$ is the occupancy of the orbital with angular momentum $l$.
To retain the structure of the Laughlin state we will demand that $U_l$ is much smaller than the bulk excitation gap (proportional to $\hat V_m$) for all $l$ that are occupied. 

We are interested in the solution to this problem in the extremely steep limit.
Precisely, we define this by
	\eqn{  \label{eq:steep} \cdots \gg U_{l+1} \gg U_l \gg U_{l-1} \gg \cdots.}
It was noted previously in Ref.~\cite{CooperSignatures} that there is a great simplification in this limit: the energy of a state is determined entirely by the largest-$l$ orbital which is occupied, and eigenstates can be constructed by successively orthogonalizing what are known as root states.
In the current paper we use this intuition to construct a fully analytic solution.

The limit we are considering in this paper, while not corresponding precisely to any experimental system that has been created thus far, is in principle possible to construct in experiment.
For both cold atom\cite{CooperReview} and qubit realisations\cite{SomethingAboutQubitFQHE} of fractional quantum Hall systems, the short-range interaction we consider is quite natural and the hard-wall confinement that we study is also possible\cite{HadzibabicWalls} (and for lattice implementations may also be natural).
 
To describe edge excitations we will need to construct a basis for symmetric polynomials by which we will multiply the Laughlin state.
A convenient basis is given by the power sums $p_n = \sum_i (z_i/R)^n$ where $R = \sqrt{2mN}$ is the radius of the Laughlin droplet.
From these we may construct a basis for all possible homogeneous symmetric polynomials via 
	\eq{P_{\bm{\lambda}} = \mbox{$\prod_{\lambda_i\in\bm{\lambda}}$} \, p_{\lambda_i}}
where ${\bm{\lambda}} =\{\lambda_1,\lambda_2, \ldots\}$ is an integer partition, made up of positive integers $\lambda_1 \geq \lambda_2 \geq \lambda_3 \ldots$.
The total degree of this polynomial is written as $|{\bm{\lambda}}| = \sum_i \lambda_i$.
When multiplied by the Laughlin wavefunction these polynomials have an orthogonality in the thermodynamic limit\cite{WenEdge, DubailEdge} given by
	\eqn{\label{eq:inner}\langle  P_{\bm{\lambda}} | P_{\bm{\mu}}\rangle  =
		\delta_{\bm{\lambda},\bm{\mu}}\,\mbox{$\prod_{n=1}^{\lambda_1}$}\,\left[(q_n!)\,(\nu n)^{q_n}\right]}
where we have written the $\bm{\lambda}$ as $\{ \lambda_1^{q_{\lambda_1}},   \ldots, 2^{q_2},  1^{q_1}\}$ to signify that the integer $n$ appears $q_n$ times.
This orthogonality, as well as the fact that corrections to Eq.~\ref{eq:inner} should be order $1/N$ smaller, is implied by conformal field theory arguments\cite{DubailEdge} and has been verified numerically \cite{DubailEdge, Us}.
These arguments and the numerics also suggest that the next sub-leading correction is of order $1/N^\frac{3}{2}$, followed by $1/N^2$ and so on.
Note that we have defined the measure of the inner product to include the Laughlin wavefunction $\langle \phi | \chi \rangle = \int \de\bm{z} \, [\phi(\bm{z})]^* \chi(\bm{z}) \, |\Psi_{m,N}(\bm{z})|^2$ where $\bm{z} = (z_1, \hdots, z_N)$ and the integral is over the complex plane $\int \frac{\de^2z}{2\pi}$ for each $z_i$.

The so-called Jack\cite{StanleyJacks} polynomials $J^{(\nu)}_{\bm{\lambda}}$ are a different set of homogeneous symmetric polynomials of degree $|\bm{\lambda}|$ which can be built from the $P_{\bm{\lambda}}$.
If the orthogonality \ref{eq:inner} were exact then the Jacks would obey
	\eqn{\label{eq:Jackorth} \langle J_{\bm{\mu}\phantom{\lambda}}^{(\nu)} |  J_{\bm{\lambda}}^{(\nu)} \rangle =  
		j_{\bm{\lambda}}(\nu) \, \delta_{\bm{\mu},\bm{\lambda}}}
where $ j_{\bm{\lambda}}(\nu)$ is a normalisation known from \cite{StanleyJacks}.
Again it has been confirmed numerically that finite size effects give order $1/N$ corrections to  Eq.~\ref{eq:Jackorth} at the first sub-leading order \cite{Us}.
The convenient feature of $J_{\bm{\lambda}}^{(\nu)}$ is that it contains a leading term with a factor of $z_i^{\lambda_1}$ and has no power of any $z_i$ higher than this.
Since the Laughlin ground state $\Psi_{N,m}$ contains particles in angular momentum orbitals up to $L_0 = m(N-1)$, this means the trial wavefunction $J_{\bm{\lambda}}^{(\nu)} \Psi_{N,m}$ contains particles in  angular momentum orbitals only up to $L_0 + \lambda_1$ and therefore its energy will be dominated by $U_{L_0+\lambda_1}$.

The main result of our paper is that in the presence of the steep confining potential \ref{eq:steep} and for large $N$, the eigenstates are given by $J_{\bm{\lambda}}^{(\nu)} \Psi_{N,m}$ with corresponding eigenenergies 
	\algn{\label{eq:Eres} E_{\bm{\lambda}} & = U_{L_0+\lambda_1}\mathcal{N}(N,\nu,\lambda_1)\prod_{j=1}^{\lambda_1}
			\frac{h_{\bm{\lambda}}(1,j)+1}{h_{\bm{\lambda}}(1,j)+\nu} \\ 
		h_{\bm{\lambda}}(1,j) & = (\lambda_1-j)\nu - 1 + \sum_{n=j}^{\lambda_1}q_n}
where we have once again written the partition in the form $\bm{\lambda}=\{\lambda_1^{q_{\lambda_1}},\hdots,2^{q_2},1^{q_1}\}$.
The proof invokes special properties of the Jacks\cite{StanleyJacks} and is outlined in the appendix with further details (and extensions of our result) given in a future paper\cite{Us}.
Corrections to these results are of order $1/N$ and of order $U_{L_0 + \lambda_1 -1}$.

The factor $\mathcal{N}(N,\nu,\lambda_1)$ has the form
	\eqn{\mathcal{N}(N,\nu,\lambda_1) = \frac{\zed_{N-1}}{\zed_N}\frac{N2^{L_0}(L_0+\lambda_1)!}{(mN)^{\lambda_1}}.}
Unfortunately, the exact form for the Laughlin normalisation $\zed_N$ is not known, so the prefactor of the energy is not fully determined in our approach (although ratios of energies are correctly obtained and are exact in the limit limit of Eq.~\ref{eq:steep} and large $N$).
However, the form can at least be partially constrained using the large-$N$ expansion in \cite{diFrancescoLaughlin}, which gives $\zed_{N-1}/\zed_N$ to be some known function of $N$ multiplied by some unknown constant which we can fit numerically.

We note that in previous works Jack Polynomials have already been used to describe quantum Hall states\cite{BernevigJacks, BernevigAnatomy,HoonJacks,ZamaereJacks,EstienneJacks,JoliceurJacks,BernevigJacks2,BernevigJacks3,EstienneJacks2, BernevigJacks4}.
However, in those works the Jack index was taken to be $-\frac{2}{m-1}$ for the Laughlin $1/m$ state, which generates a polynomial describing the \textit{full} state, including the Laughlin factor.
Here we use the Jacks with index $1/m$ only as the polynomials multiplying the Laughlin.
One might wonder whether these polynomials are not simply an alternative expression of the full Jack states used previously.
However, were one to extract the Laughlin factor from those full states they would find\cite{BarattaJacks} the Jack in front to have index $\frac{2}{m+1}$, which is not orthogonal under the Laughlin measure.

\begin{figure}[h!]
\centering
\includegraphics[trim=0.1cm 0cm 0.1cm 0.5cm, width=0.5\textwidth]{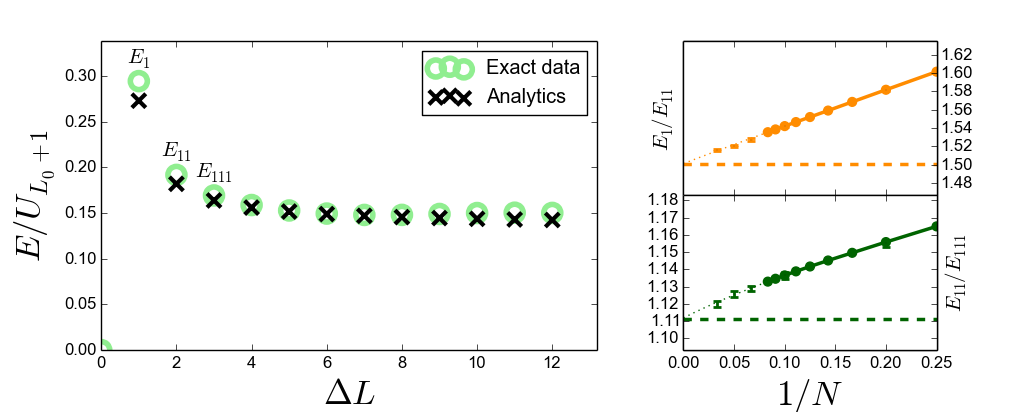}
\caption{A comparison of exact diagonalization data and our analytic result for filling fraction $\nu=1/2$ and system size $N=12$ in the $\lambda_1=1$ branch. For the analytics we fit $\zed_{N-1}/\zed_N$ using this data (and check the form against Monte Carlo integration). The right panel demonstrates the convergence of ratios of certain energies (which should be exact without the problematic $\zed_{N-1}/\zed_N$ factors). Circles represent exact diagonalization data and dumbbells represent Monte Carlo data with a size corresponding to the total error in the result. The dotted line is an extrapolation to $N\to\infty$ which uses the first two sub-leading corrections (i.e, we fit to some function $a+b/N+c/N^\frac{3}{2}$). The horizontal dashed lines are the `exact' predictions for the thermodynamic limit behaviour.}
\label{exact_compare1}
\end{figure}

\begin{figure}[h!]
\centering
\includegraphics[trim=0cm 0cm 0cm 0cm, width=0.5\textwidth]{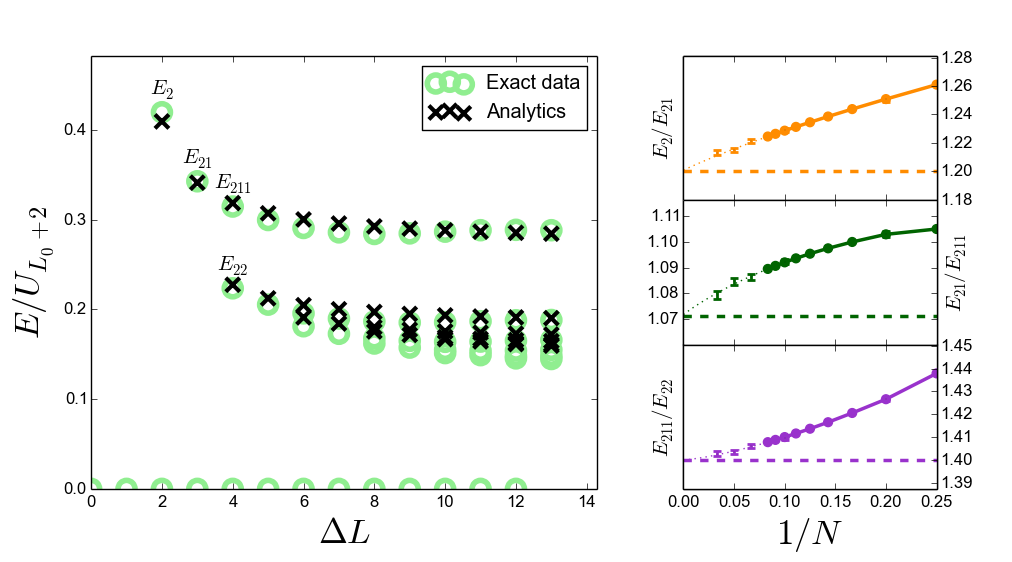}
\caption{A comparison of exact diagonalization and analytics for the $\lambda_1=2$ band for $\nu=\frac{1}{2}$ and $N=12$ alongside the finite size scaling of energy ratios. Once again, the $N\to\infty$ extrapolation is a fit of the first two sub-leading corrections. Here the $1/N^\frac{3}{2}$ terms are large enough to be important.}
\label{exact_compare2}
\end{figure}

We now turn to consider the physics of our result.
First, we note that the spectrum breaks into branches having different values of $\lambda_1$, with the corresponding piece having energy proportional to $U_{L_0 + \lambda_1}$.
The ratio of energies between the branches is determined by the structure of the confining potential $\hat U$, and is non-universal.
Secondly within each branch, we find that the the highest energy corresponds to the {\it smallest} angular momentum (we will describe a case of this in detail below and describe the physics of this) in strong contrast to the traditional Luttinger liquid picture.
Nor do the energies bear any resemblance to that of the Calogero-Sutherland model.
In Figs.~\ref{exact_compare1} and \ref{exact_compare2} we show the lowest two branches of excitation spectrum for $\nu=1/2$ and 12 particles.
Despite the small system size (and putative $1/N$ corrections) our analytic results are in almost perfect agreement with exact diagonalization.

For example, the lowest branch of excitations (given in the figure) correspond to the partitions  $\bm{\lambda} = \{1^{\Delta L}\}$.
Here, the maximum amount of angular momentum added to any one electron is 1, giving an energy spectrum of the form
$E_{1^{\Delta L}} = U_{L_0+1}\mathcal{N}(N,\nu)\frac{\Delta L}{\Delta L + \nu - 1}$ which is shown in Fig.~\ref{exact_compare1}.
The second branch has modes at energy scale $U_{L_0+2}$ and partitions of the form $\{2^a, 1^b \}$ with $a>0$ and $b \geq 0$ where $\Delta L = 2a + b$ as shown in Fig.~\ref{exact_compare2}. 

As shown in the figures for each branch, the energy is largest for the smallest angular momentum.
To understand this we consider first the case of $\nu=1$ which is noninteracting fermions.
The ground state is simply a single slater determinant of all orbitals filled up to angular momentum $L_0$.
For the lowest branch of excitations (the $U_{L_0 + 1}$ branch) one electron is pushed into the angular momentum $L_0+1$ orbital, leaving a single hole in one of the previous filled orbitals (and no electrons are promoted to orbitals with higher angular momentum than $L_0+1$).
However, given the steep edge condition Eq.~\ref{eq:steep}, the occupancy of any orbitals other than the $L_0 + 1$ orbital has no bearing on the energy, and thus the energy of the excitation is independent of the angular momentum of this hole.
Thus the $U_{L_0+1}$ branch of excitation has energy independent of $\Delta L$.  

In the case of $\nu=1/m$ with $m>1$ the physics is somewhat similar.
If one were to crudely think of the system as being made of composite fermions filling orbitals\cite{JainComposite}, one might guess that the energy is similarly independent of the angular momentum of the composite fermion hole (again so long as we only insert a single hole, no electrons are promoted to orbitals with higher angular momentum than $L_0+1$).
However, one should remember that electron orbitals (for $L \leq L_0+1$) are neither completely filled nor completely empty, and it is, in fact, the density of electrons in the $L_0 + 1$ orbital which completely determines the energy of the state.
In this case, the orbital $L_0 + 1$ becomes {\it less} occupied when there are other electrons near to it, because of the correlation holes (Jastrow factors) carried by the other electrons.
Thus the energy is lowest when the holes are pushed further from the edge --- i.e., when the angular momentum is the largest. 

The methods introduced in this paper can be extended in several ways.
For example, with more analytic effort corrections to some of the results may be obtained which keep sub-leading terms in orders of $U_{l-1}/U_l$.
The technique may also be applied to the bosonic edge excitations of the other quantum Hall states, such as the Moore-Read state, although analysis of non-bosonic edge excitations is not as obviously tractable.
Finally, it is interesting to think about how $1/N$ corrections to Eq.~\ref{eq:inner} might give finite sized corrections to our results.

In summary we have given an analytic solution for the edge excitations of Laughlin states in a steep confinement limit.
We find that eigenfunctions are given by the Jack polynomials analogous to the Calogero-Sutherland model, but the eigenenergies differ from that model.

{\bf Appendix:}  We want to evaluate 
	\eq{\hat U_{\bm{\mu},\bm{\lambda}} = \frac{\langle J_{\bm{\mu}}^{(\nu)}|\hat U|J_{\bm{\lambda}}^{(\nu)}\rangle}
		{\sqrt{j_{\bm{\mu}}(\nu)j_{\bm{\lambda}}(\nu)}}}
which we claim has the form $\hat U_{\bm{\mu},\bm{\lambda}} = E_{\bm{\lambda}} \delta_{\bm{\mu},\bm{\lambda}}$.   Since $\hat U$ is rotationally invariant, polynomials with different $|{\bm{\lambda}}|$ (i.e., different total angular momentum) must be orthogonal.   The nontrivial situation is when $|\bm{\mu}| = |\bm{\lambda}|$.

The key simplification stems from the steep edge condition, \ref{eq:steep}.  To leading order, we need only identify the highest angular momentum orbital which is occupied, say $z^l$, and we can focus only on the potential from this orbital $U_l$.   Given the structure of the Jacks this will be for $l=L_0 + \lambda_1$.    We introduce notation from \cite{StanleyJacks} and use $[z_k^n]$ to mean that we are isolating the coefficient of  $z_k^n$ and throwing away terms that do not have this factor (we will assume that this operator applies only to the holomorphic part of the wavefunction and ignores the Gaussian factor $e^{-|z|^2/4}$ which can be thought of as part of the measure). Applying this factorisation to the wavefunction $J_{\bm{\lambda}}^{(\nu)} \Psi_{m,N}$, we have obviously
	\eq{[z_k^{L_0+\lambda_1}] \, J_{\bm{\lambda}}^{(\nu)} \Psi_{m,N}  =  \left([z_k^{\lambda_1}]  J_{\bm{\lambda}}^{(\nu)} \right)
		\left( [z_k^{L_0}] \Psi_{m,N} \right)}
since the only way we can factorise out $z_i^{L_0+\lambda_1}$ is by saturating the angular momentum contributions from both the Jack polynomial and the Jastrow factors. We now invoke identity 5.2 in \cite{StanleyJacks} which states
	\eq{[z_k^{\lambda_1}] J_{\bm{\lambda}}^{(\nu)}({\bf z}) = (z_k/R)^{\lambda_1} d_{\bm{\lambda}}
		(\nu)J_{\bm{\lambda}^-}^{(\nu)}({\bf z}_{\cancel{k}})}
Note that $z_k$ is normalised by $R$ because our Jacks are composed of the $P_{\bm{\lambda}}$, which are defined as sums of $(z_k/R)^n$.  Here we have defined the notation ${\bf z}_{\cancel{k}}  = \{z_1, \ldots, z_{k-1}, z_{k+1}, \ldots, z_N\}$ and a new partition, $\bm{\lambda}^-$, which is our original partition with the first element removed, $\bm{\lambda}^-=\{\lambda_2,\lambda_3,\hdots\}$. The function $d_{\bm{\lambda}}(\nu)$ is given in \cite{StanleyJacks}. Similarly, we can apply the factorisation to the Jastrow factors to find
	\eq{[z_k^{L_0}]\prod_{i<j}(z_i-z_j)^m = z_k^{L_0}(-1)^{m(k-1)}\prod_{i<j\neq k}(z_i-z_j)^m.}
We thus have
	\alg{&\hat U[J_{\bm{\lambda}}^{(\nu)}\Psi_{m,N}] = 
		U_{L_0 + \lambda_1} (\zed_{N-1}/\zed_{N})^{1/2} d_{\bm{\lambda}}(\nu) \\
		& \sum_k \left(\frac{z_k}{R}\right)^{\lambda_1} J_{\bm{\lambda}^-}^{(\nu)}({\bf z}_{\cancel{k}})
			(-1)^{m(k-1)}z_k^{L_0}   \Psi_{m,N-1}({\bf z}_{\cancel{k}}) e^{-|z_k|^2/4}}
with sub-leading corrections of order $U_{L_0+\lambda_1-1}$ and of order $1/N$.  With the same manipulation on the bra  $J_{\bm{\mu}}^{(\nu)}$, we obtain the inner product
	\alg{&\hat U_{\bm{\mu},\bm{\lambda}} = U_{L_0+\lambda_1}\frac{d_{\bm{\mu}}(\nu)d_{\bm{\lambda}}(\nu)}
		{\sqrt{ j_{\bm{\mu}}(\nu)j_{\bm{\lambda}}(\nu)} R^{2\lambda_1}} \frac{\zed_{N-1}}{\zed_N}
		\langle J_{\bm{\mu}^-}^{(\nu)}|J_{\bm{\lambda}^-}^{(\nu)}\rangle \\
		& \qquad \sum_k \int \frac{\de^2 z_k}{2\pi} (z_k^*)^{L_0 + \mu_1}(z_k)^{L_0+\lambda_1} e^{-|z_k|^2/2} \\
		& = U_{L_0+\lambda_1}\left[\frac{2^{L_0}(L_0+\lambda_1)!}{m^{\lambda_1} N^{\lambda_1-1}}\frac{\zed_{N-1}}{\zed_N}
		\right] \frac{j_{\bm{\lambda}^-}(\nu)}{j_{\bm{\lambda}}(\nu)}(d_{\bm{\lambda}}(\nu))^2 \delta_{\bm{\lambda},{\bm{\mu}}}}
which is diagonal as claimed.   Using results from Ref.\cite{StanleyJacks} for the form of $j_{\bm{\lambda}}$ and $d_{\bm{\lambda}}$, with a bit of algebra, we simplify the result to that given in Eq.~\ref{eq:Eres}.

{\bf Acknowledgments:} We are very grateful to A. Lamacraft and R. Bondesan for enlightening discussions and useful comments.  This work was supported by EPSRC  grants
Nos. EP/I032487/1, EP/I031014/1, and EP/N01930X/1.   We would also like to acknowledge the DiagHam package, maintained by N. Regnault, which provided key assistance to this work.  Statement of compliance with EPSRC policy framework
on research data: This publication is theoretical
work that does not require supporting research data.

\bibliography{bib}{}

\end{document}